\documentclass{ws-procs975x65}
\begin{document}

\title{Information-theoretic corrections to black hole area quantisation}

\author{Rajesh R. Parwani}

\address{Department of Physics, National University of Singapore,\\
Kent Ridge, 117542, Singapore\\
E-mail: parwani@nus.edu.sg}

\begin{abstract}
Nonlinear corrections are proposed to the discrete equispaced area spectrum of quantum black holes obtained previously in some quantisation schemes. It is speculated that such a modified spectrum might be related to the fine structure found using the loop quantum gravity approach. 
\end{abstract}


\bodymatter

\section*{Summary}

A long time ago, Beckenstein and Mukhanov \cite{becka,beckb} conjectured that the area spectrum of quantum black holes should be discrete and equispaced. This was then confirmed by several authors using different approaches. For example Barvinsky et.al. \cite{bar} performed a collective coordinate quantisation of a classical black hole, obtaining a Schrodinger eigenvalue equation for the black hole area spectrum,
\begin{equation}
\hat{A} \psi = a_n \psi \, . \label{lab}
\end{equation}
Since the area operator $\hat{A}$ has the form of the Hamiltonian for a one dimensional simple harmonic oscillator, $\hat{A} = X^2 + \Pi^2$, the spectrum is discrete and equispaced. 

The approach of Ref.\cite{bar} has the advantage of being quite general, independent of any microscopic model of the quantum black hole, see further discussion and references in Ref.\cite{das}. But one may then wonder if the result  represents an approximate picture, with some fine structure having been averaged over. 

Let us investigate potential corrections to the above results using information-theoretic arguments. It is known \cite{Scha,Schb} that the structure of the usual Schrodinger equation may be motivated, starting from classical ensemble dynamics, using the maximum uncertainty (entropy) method \cite{Jaynesa,Jaynesb}. The maximum uncertainty approach is used in many disciplines; its basic philosophy being to minimise any bias (maximise uncertainty) when choosing probability distributions. In this approach the uncertainty is quantified by a measure which can be derived axiomatically; the simplest such measure for the present context has the form of the Fisher measure \cite{Sch2}.

Deformations of the Fisher measure then lead to nonlinear generalisations of the Schrodinger equation which presumably model effects of shorter distance physics. In Ref.\cite{RP1} it was argued that the Kullback-Liebler measure was the most natural generalisation of the Fisher measure and the corresponding nonlinear  Schrodinger equation constructed. Mathematical properties of that equation have been investigated in several papers \cite{Tabia,RP2a,RP2b,RP2c} and it has also been used in the context of quantum cosmology where it was shown that the nonlinearity can help avoid Big Bang singularities in some simple cosmologies, replacing them with a bounce for the evolution of the universe \cite{Nguyen}.

Using the nonlinear equation of Ref.\cite{RP1} one may determine corrections to the area spectrum (\ref{lab}). For weak nonlinearity, the perturbative result is \cite{Tabia}  
\begin{equation}
\delta A \propto  \sqrt{\eta(1 - \eta)}\ \
(1-4 \eta) \ n^{1.41} \label{EPX}
\end{equation}
where $n$ is the simple harmonic oscillator quantum number and $0<\eta<1$ a free parameter. 

How may one test the above suggested correction? Well, the area of a quantum black hole has been investigated using the methods of loop quantum gravity (LQG). In this nonperturbative approach to quantum gravity, discreteness arises in the area and volume elements of space. Although the area spectrum is discrete, it is not equidistant, but the degeneracies are clustered and the spectrum is effectively equidistant \cite{agu}. Since the LQG approach seems to be revealing a fine structure not seen using the approach of Ref.\cite{bar}, one may speculate that the difference is due, to leading order, to the correction (\ref{EPX}): That is, the distance between the peaks of the clusters deviates from equidistance by an amount given by Eq.(\ref{EPX}). In principle one may test this suggestion using the data of Ref.\cite{agu}  but this has not yet been done.

If one further assumes that the entropy of the black hole is still proportional to its area, even with the perturbative correction (\ref{EPX}), then the entropy of the black hole would be discrete but not equispaced,
$S = a + bn + cn^{1.41}$ with the constant $c$ dependent on $ 0< \eta <1$.
 Indeed in Ref.\cite{cor} the entropy of black holes was calculated within LQG  by counting quantum microstates and it was found to have an effective band structure, with also an approximately equidistant spacing. However it was noticed that, for small black holes at least, the change in entropy increased as the black holes grew \cite{cor}. 
 
Note that one may test the $n$ dependence in (\ref{EPX}) eventhough the parameter $\eta$ would need to be adjusted to fit the data. Furthermore one may differentiate the information-theoretically motivated nonlinearity from  other nonlinearities because of the different $n$ dependence they predict \cite{Tabia}.

In summary, an information-theoretic framework has been used to model the new physics that is generally believed to exist in the quantum gravity regime. This approach was previously used to study the evolution of cosmologies near the Big Bang \cite{Nguyen}. In the present paper a similar approach has been  proposed for black holes. There are some potentially interesting quantitative comparisons that remain to be performed between this approach and that of loop quantum gravity. It would also be useful to obtain non-perturbative corrections to the spectrum (\ref{lab}) due to the proposed nonlinearity, going beyond the perturbative calculation of Ref.\cite{Tabia} that yielded (\ref{EPX}).

\section*{Acknowldegement} 
I thank the organisers for the opportunity to present this work and to Saurya Das for helpful correspondence.

\end{document}